\documentstyle[preprint,aps,epsf]{revtex}
\begin{document}

\title{Statistical Fluctuations of  Electromagnetic Transition Intensities and Electromagnetic Moments in 
$pf$-Shell Nuclei}

\author{A. Hamoudi,$^1$ R.G. Nazmitdinov,$^{1,2}$ E. Shahaliev,$^1$  
and Y. Alhassid$^3$}
\address{$^1$Bogoliubov Laboratory of Theoretical Physics,
Joint Institute for Nuclear Research, 141980 Dubna, Russia}
\address{$^2$Max-Planck-Institut f\"ur Physik komplexer
Systeme, D-01187 Dresden, Germany}
\address{$^3$Center for Theoretical Physics,
Sloane Physics Laboratory, Yale University, New Haven,
Connecticut 06520,  USA}
\date{\today}
\maketitle

\begin{abstract}

We study the fluctuation properties of $\Delta T=0$ electromagnetic transition intensities 
and electromagnetic moments in  $A \sim 60$ nuclei within the framework of the interacting shell model, using a realistic effective interaction for
 $pf$-shell nuclei with a $^{56}$Ni  core. The distributions of the transition intensities and of the electromagnetic moments are well described by the Gaussian orthogonal ensemble of random matrices. In particular, the transition intensity distributions follow a Porter-Thomas distribution. When diagonal matrix elements (i.e., moments) are included in the analysis of transition intensities, we find that the distributions remain Porter-Thomas except for the isoscalar $M1$. The latter deviation is explained in terms of the structure of the isoscalar $M1$ operator.

\end{abstract}

\draft

\narrowtext

 Random matrix theory (RMT) \cite{Meh} was originally introduced to explain
 the statistical fluctuations of neutron resonances  in compound nuclei 
\cite{Por}.
  The theory assumes that the nuclear Hamiltonian belongs to an ensemble
 of random matrices that are consistent
 with the fundamental symmetries of the system. In particular, since the
 nuclear interaction preserves time-reversal symmetry, the relevant ensemble
 is the Gaussian orthogonal ensemble (GOE). The use of RMT in the compound 
nucleus was justified by the complexity of the nuclear system.  Bohigas 
 {\it et al} \cite{Boh} conjectured that  RMT describes the statistical 
fluctuations of a quantum system whose associated classical dynamics is 
chaotic.
 RMT has become a standard tool for analyzing the universal statistical 
fluctuations in chaotic systems \cite{Br,Guh,Al}.

 The  chaotic nature of the single-particle dynamics  in the nucleus can be studied in the mean-field approximation.  The interplay between shell structure and fluctuations in the single-particle spectrum has been understood in terms of the classical dynamics of the nucleons in the corresponding deformed mean-field potential \cite{H94,H99}.
However, the residual nuclear interaction mixes different mean-field configurations and affects the statistical fluctuations of the many-particle spectrum and wave functions. 
These fluctuations can be studied using various nuclear structure models. The statistics of the
 low-lying collective part of the nuclear spectrum have been studied in the framework of the interacting boson model \cite{Al1,Al2}, in which
 the nuclear fermionic space is mapped onto a much smaller space of  bosonic degrees of freedom. Because of the relatively
small number of degrees of freedom in this model, it was also possible to
relate the statistics to the underlying  mean-field collective
dynamics.  At higher excitations, additional degrees of freedom (such as broken pairs) become important \cite{Al4}, and
 the effects of interactions on
the statistics must be studied in larger model spaces.
The interacting shell model offers an attractive framework for such studies, in which realistic effective interactions are available and where the basis states are labeled by  exact quantum numbers of angular momentum, isospin and parity \cite{Zel}.

 RMT makes definite predictions for the statistical fluctuations of both 
the eigenfunctions and the spectrum.  The  electromagnetic transition 
intensities in a nucleus are observables that are sensitive to the wave functions, and the study
of their statistical distributions should complement  \cite{Al1,Al2}
 the more common spectral analysis.
Using the interacting shell model, $B(M1)$ and $B(E2)$ transitions 
were recently analyzed in $^{22}$Na \cite{m1}
 and found to follow the Porter-Thomas distribution
\cite{PT}, in agreement with RMT and with the previous finding 
of a Gaussian distribution for the eigenvector components \cite{v1}-\cite{v4}.

 Most  studies of statistical fluctuations in the shell model have been restricted to lighter nuclei ($A \alt 40$) where complete 
$0 \hbar \omega$ calculations
 are feasible (e.g., $sd$-shell nuclei). 
Here we study the fluctuation properties
of $\Delta T=0$ electromagnetic transition intensities in nuclei with $A \sim 60$. We also study the statistics of the diagonal matrix elements, which describe the corresponding electric or magnetic moments.  We find that the statistics of both the transition intensities and the electromagnetic moments are well described by RMT.  In RMT, off-diagonal matrix elements have a Gaussian distribution with zero mean, leading to a Porter-Thomas distribution of the transition intensities. Diagonal elements also have a Gaussian distribution but with non-zero mean. Thus the squares of the diagonal matrix elements should also follow a Porter-Thomas distribution but only after the mean value is subtracted from the diagonal elements.   Experimentally it is difficult to measure the moments of excited states, while transition intensities are more readily accessible. 

When the good quantum numbers (isospin, spin and parity) of the initial and final states are the same, we can include the diagonal matrix elements in the analysis of the transition intensity distribution. The number of diagonal elements is much smaller than the number of off-diagonal elements, and generically we do not expect the diagonal elements to modify the Porter-Thomas statistics of the transition intensities. This is confirmed in our shell model calculations with the exception of the 
 $ M1$ operator in self-conjugate nuclei ($T_z=0$), where  
a significant deviation from a Porter-Thomas distribution is found.\footnote{We find a similar deviation in $sd$-shell nuclei.} 
 This deviation is explained in terms of the special structure of the isoscalar $M1$ operator, which leads to diagonal matrix elements that are much larger than its off-diagonal elements.
 
Full $pf$-shell calculations  are currently not feasible
for $A \sim 60$ nuclei. We therefore perform  $pf$-shell calculations  with  $^{56}$Ni as a core, i.e., we  assume fully occupied $f_{7/2}$ orbits and consider all possible many-nucleon configurations defined by the $0f_{5/2}$, $1p_{3/2}$ and $1p_{1/2}$ orbitals. The effective interaction is chosen to be the isospin-conserving F5P interaction \cite{in1}.
This interaction is successful in describing the mass range $A \sim 57-68$. The calculations were performed using the shell model program 
OXBASH \cite{2}. 

\begin{figure}

\epsfxsize=12 cm 
\centerline{\epsffile{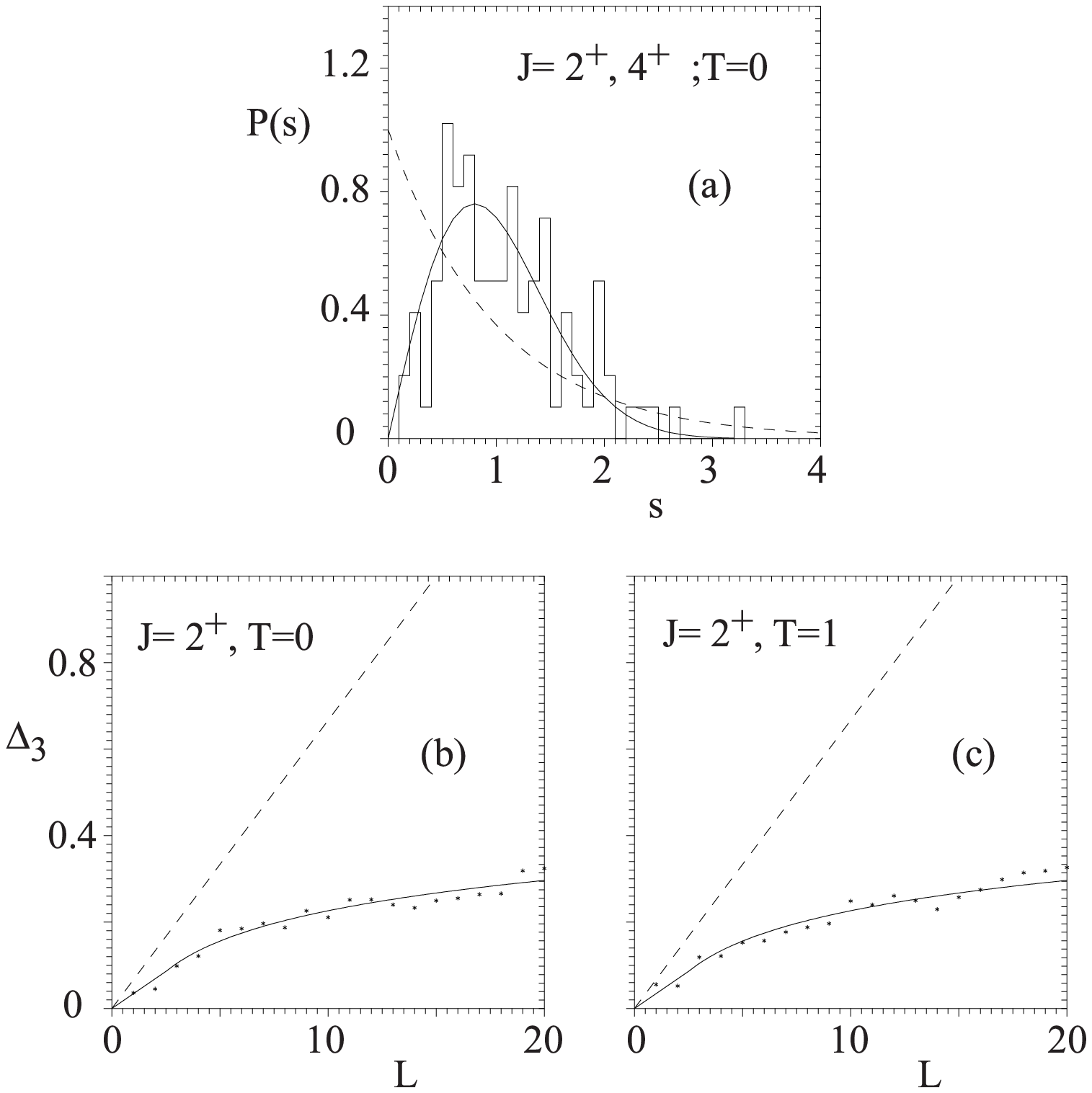}}

\vspace{5 mm}

\caption
{ Spectral statistics in $A=60$ nuclei.  (a) Nearest-neighbor level spacing
 distribution $P(s)$ for $T=0$ states with $J=2,4$ (histograms). The solid
line is the Wigner-Dyson distribution and the dashed line is the Poisson
distribution. Bottom: $\Delta_3$ statistic for $J=2$ levels with (b) $T=0$ and (c) $T=1$. Also shown are the GOE (solid lines) and Poisson  (dashed lines) limits.}
\label{fig1}

\newpage

\epsfxsize=12 cm 
\centerline{\epsffile{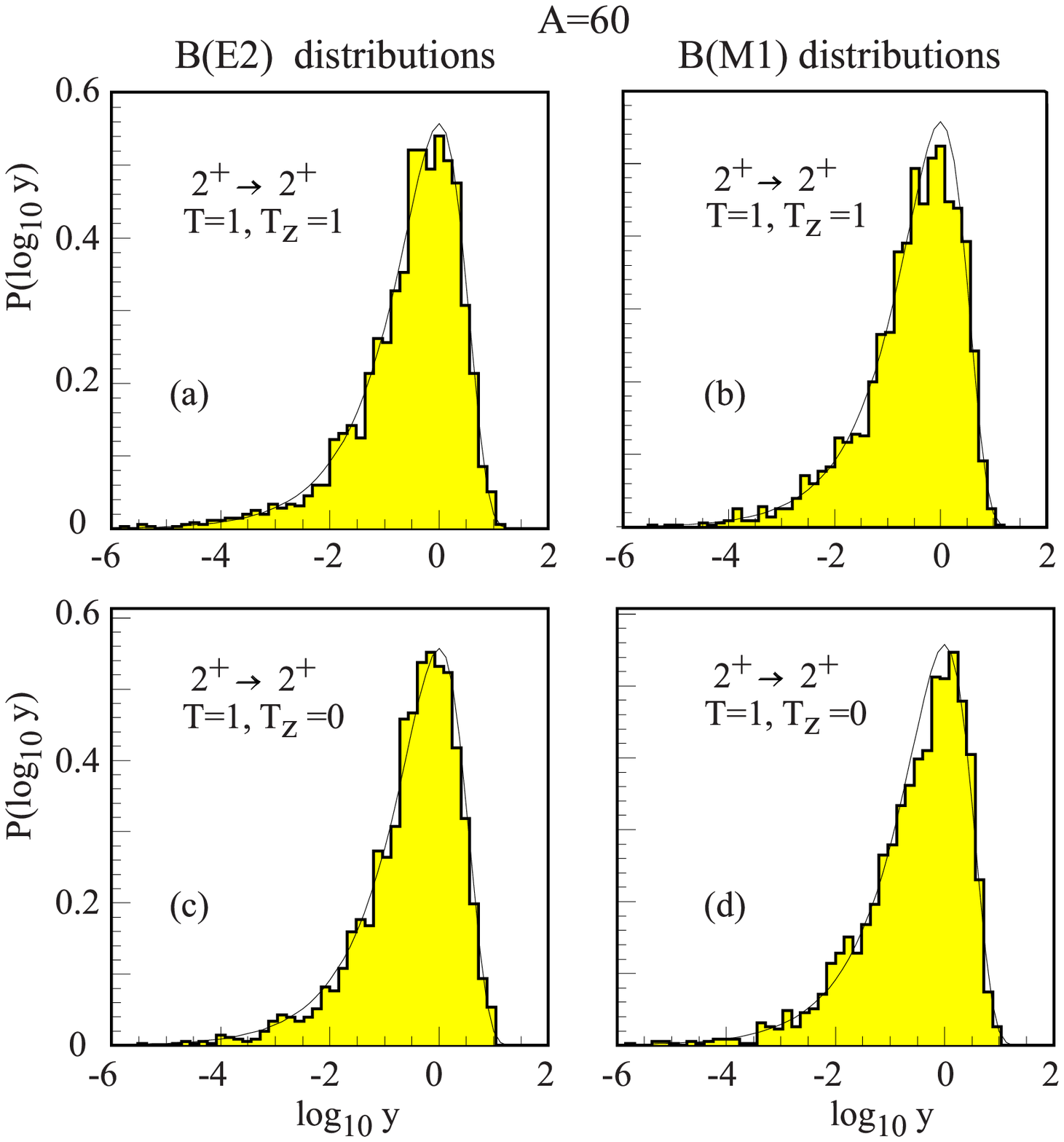}}

\vspace{5 mm}

 \caption
{ The  $B(E2)$  (left panels) and $B(M1)$ (right panels) intensity 
distributions (histograms) for the
 $2^{+}, T =1 \rightarrow 2^{+}, T =1$ transitions  in $A= 60$ nuclei: (a,b) $T_z=1$; (c,d) $T_z=0$.  The solid  lines describe the Porter-Thomas distribution
(Eq. (\protect\ref{chi2-nu}) with $\nu=1$).}
\label{fig2}

\newpage

\epsfxsize=12 cm 
\centerline{\epsffile{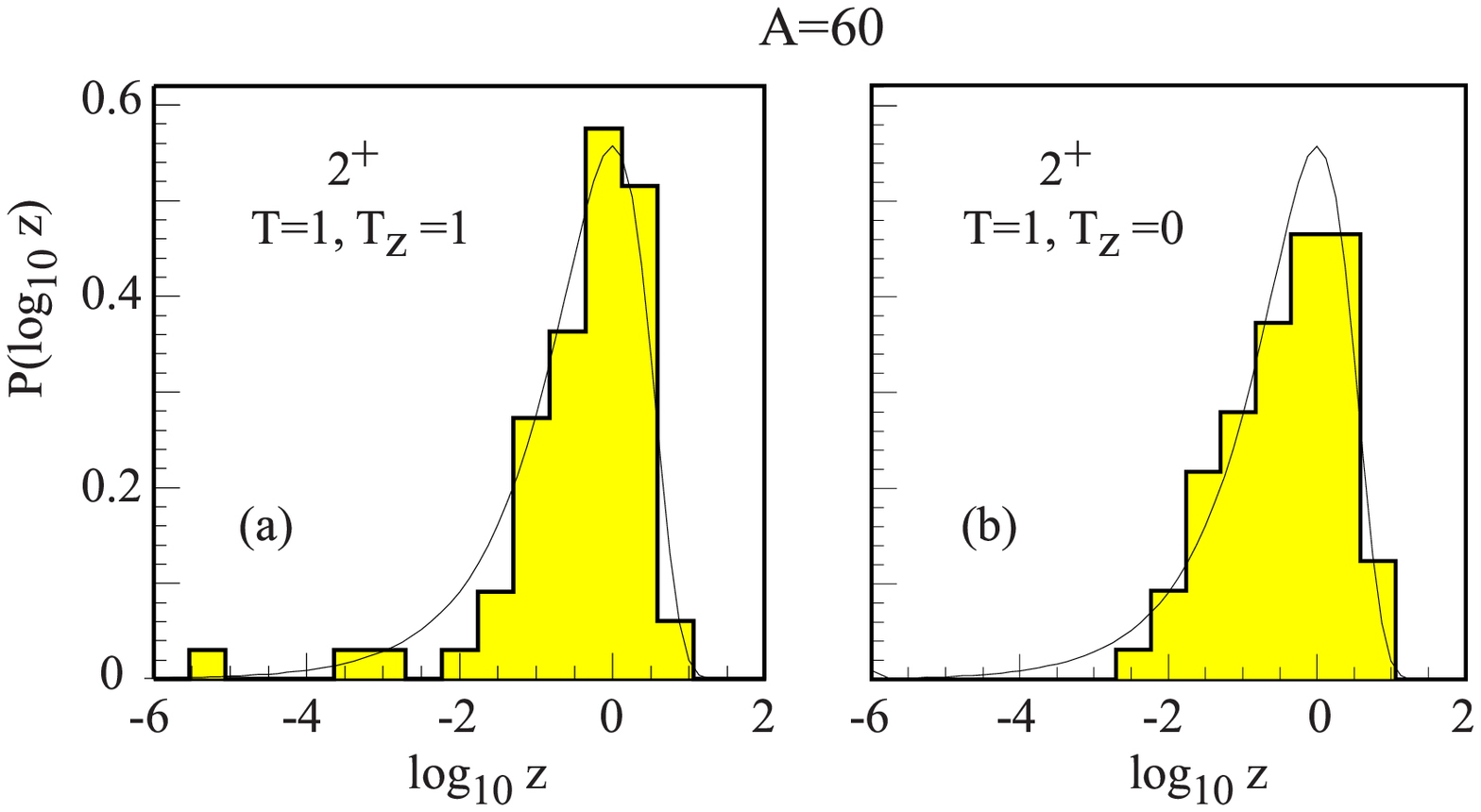}}

\vspace{5 mm}

\caption
{Statistics of the $M1$ magnetic  moments of the $J=2^+,T=1$ states in $A=60$ nuclei: (a) $T_z=1$; (b) $T_z=0$. The histograms show the distribution $P(\log_{10}z)$ where $z$ are the squares of the $M1$ diagonal reduced matrix elements whose mean value (\protect\ref{diag-mean}) has been subtracted.   The solid lines describe Porter-Thomas distributions.}
\label{fig3}

\newpage

\epsfxsize=12 cm 
\centerline{\epsffile{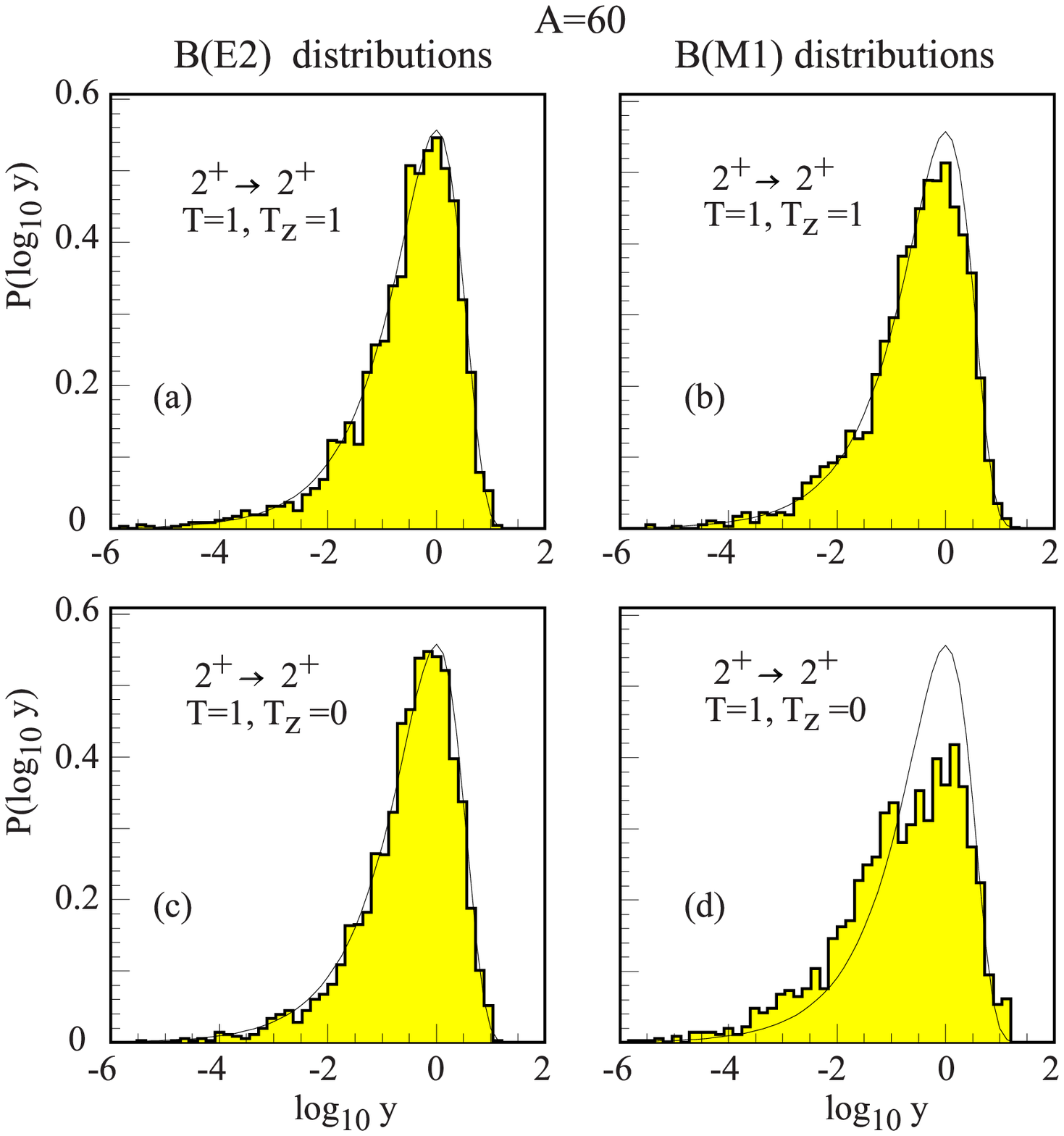}}

\vspace{5 mm}

\caption
{$B(E2)$  (left panels) and $B(M1)$ (right panels) intensity 
distributions (histograms) for $A=60$ nuclei when diagonal matrix elements are included in the analysis. Shown are the same cases as in Fig. \protect\ref{fig2}. Notice that the $B(M1)$ distribution deviates from the 
Porter-Thomas distribution in the self-conjugate $T_z=0$ nucleus.}
\label{fig4}

\end{figure}

   To test the validity of RMT in the above model space, we first study the spectral fluctuations for states with good spin and isospin 
(all states in the $pf$-shell model space have the same parity).    
Fig. \ref{fig1}(a)
 shows the nearest-neighbors level spacing distribution $P(s)$ for the 
unfolded $T=0$ levels. To improve the statistics we combine the spacings of the $J=2$ states and the $J=4$ states.
 The calculated distribution (histograms)
is in agreement with the Wigner-Dyson distribution (solid line), and  level repulsion  is clearly observed at small spacings. The Poisson distribution, 
which  corresponds to a random sequence of levels (and characterizes regular systems), is shown by the dashed line for comparison.  Another measure of fluctuations is the spectral rigidity described by  Dyson's  $\Delta_3$ statistic \cite{Meh}. 
The bottom panels of Fig. \ref{fig1} show $\Delta_3(L)$ versus  $L$  for 
$J=2$ states with $T=0$ [panel (b)] and $T=1$ [panel (c)].  The results agree well with the GOE limit (solid lines).

We now consider the electromagnetic transition intensities.
 Denoting by $B(\overline{\omega}L; i \to f)$ the reduced transition 
probability from an initial state $|i \rangle$ to a final state $|f \rangle$, with $\bar{\omega}$ indicating the electric ($E$) or magnetic ($M$) character of the transition, and $2^{L}$ the multipolarity, we have \cite{Bru}
\begin{equation}
\label{bel}
B(\overline{\omega}L; J_i T_i T_z \rightarrow J_f T_f T_z)=
\frac{|\delta_{T_i T_f}M_{is}(\overline{\omega}L)
-  (T_i T_z 10|T_f T_z ) M_{iv}(\overline{\omega}L)|^2} {(2J_i+1)(2T_i+1)} \;.
\end{equation}
Here  $M_{is}(\overline{\omega}L)$
and $M_{iv}(\overline{\omega}L)$ are the triply reduced matrix elements
 for the isoscalar and isovector components of  the transition operator,
 respectively.  Note that these matrix elements depend on $J_i, T_i$
 and $J_f, T_f$ but not on $T_z$.
 For $\Delta T=0$ transitions ($T_i=T_f =T$) the
isospin Clebsch-Gordan coefficient in Eq. (\ref{bel})  is simply given by
\begin{equation}
(T T_z 10|T T_z )={T_z}/\sqrt{T(T+1)} \;.
\end{equation}
 It follows that  the isovector component  in Eq. (\ref{bel}) is absent
 for self-conjugate  (i.e. $T_z=0$) nuclei. Thus the statistics of the
 isoscalar component
of an electromagnetic transition operator can be inferred
directly from the study of $\Delta T=0$ transitions in $T_z=0$ nuclei.  
 Consequently, we
 are able to test the sensitivity of the statistics to
the isovector and the isoscalar contributions. Below we present results for
 E2 and M1 transitions.   The $E2$ transitions were calculated using
standard effective charges of  $e_p=1.5$
and $e_n=0.5$. We note, however, that  the isoscalar and isovector
 components of $E2$ are  (up to a proportionality constant) independent
 of the effective charges, and their corresponding statistics are thus also
 independent of  the particular choice of effective charges.

To study the fluctuation properties of
the transition rates, it is necessary to divide out any secular variation
 of the average strength function versus the initial and final
energies.  We do this by applying the method of Refs. \cite{Al3} and  \cite{Al2}. We calculate  an average transition
strength at an initial energy $E$ and final energy $E^\prime$ from 
\begin{eqnarray}
\label{av1}
\langle B(\overline{\omega}L; E,E^\prime) \rangle=\frac{\sum_{i,f}B(\overline{\omega}L; i
\rightarrow f) e^{-(E-E_i)^2/2\gamma^2 }
e^{-(E^\prime-E_f)^2/2\gamma^2}}{\sum_{i,f}e^{-(E-E_i)^2/2\gamma^2}
e^{-(E^\prime-E_f)^2/2\gamma^2}} \;,
\end{eqnarray}
where $\gamma$ is a parameter chosen as described below.
 For fixed values of  the initial ($J_i,T$) and final ($J_f,T$), we calculate 
from Eq. (\ref{bel})  the intensities
$B(\overline{\omega}L; i\rightarrow f)$. All transitions  of a given operator
(e.g., $M1$ or $E2$) between the initial and final states of the  given
spin and isospin classes have been included in the statistics.
We remark that the energy levels used in (\ref{av1}) are the unfolded energy
levels\cite{BG} characterized by a constant mean spacing.
The value of $\gamma$ in (\ref{av1}) has been chosen to be large enough
to minimize effects arising from
the local fluctuations in the transition strength but not
so large  as  to wash away the secular energy variation of the average
intensity.  In the present calculations we use $\gamma \sim 2.5$. 
 We renormalize the actual intensities by dividing out their smooth part
\begin{equation}\label{av2}
y_{fi}= {B(\overline{\omega}L;  J_i T T_z  \to J_f T T_z) \over
\langle B(\overline{\omega}L; E,E^\prime) \rangle} \;,
\end{equation}
and  construct their distribution using  bins that are equally spaced in $\log_{10} y$.  The choice of  $\log_{10}y$ as the variable allows us to display the distribution of the weak transitions over several orders of magnitude \cite{Lev}. 

  In RMT we expect a Porter-Thomas 
distribution for $P(y)$, i.e., a $\chi^2$ distribution 
 distribution in $\nu =1$ degrees of freedom \cite{Lev}. A $\chi^2$ 
distribution in $\nu$ degrees of freedom is given by
\begin{equation}\label{chi2-nu}
P_{\nu}(y)=(\nu /2<y>)^{\nu /2}y^{\nu /2 - 1}e^{-\nu y/2<y>}/\Gamma(\nu/2) \;.
\end{equation}
 Indeed, consider a transition operator ${\cal T}$ and its matrix elements $\langle f | {\cal T} | i \rangle$ between an initial state $|i \rangle$ and a final state $|f\rangle$. Using a complete (and fixed) set of states $|m \rangle$, we have
\begin{equation}\label{transition-element}
 \langle f | {\cal T} | i \rangle = \sum\limits_{m,n} {\cal T}_{m,n} \psi^\ast_f(m) \psi_i(n) \;,
\end{equation}
where $\psi_i(n) \equiv \langle n | i \rangle$ represents the eigenfunction $|i\rangle$ in the fixed chosen basis and $T_{m,n}=\langle m|T |n\rangle$. In RMT, $\psi_i(n)$ are Gaussian random variables, and by the central limit theorem it follows from (\ref{transition-element}) that $\langle f |{\cal T}| i\rangle$ is also Gaussian. In RMT, two different eigenfunctions are uncorrelated $\overline{\psi^\ast_f(m) \psi_i(n) } =0$, and consequently $\overline{\langle f | {\cal T} | i \rangle}=0$. Similarly, using Wick's theorem for Gaussian variables, the variance is given by
\begin{equation}\label{tran-variance}
\sigma^2(\langle f | {\cal T} | i \rangle) = \sum\limits_{m,n,m',n'} {\cal T}^*_{m,n} {\cal T}_{m',n'} \overline{\psi_f(m) \psi_f^*(m')} \;\; \overline{\psi_i(n) \psi_i^*(n')} ={1 \over N_i N_f} {\rm Tr}\; \left(P_i {\cal T}^\dagger P_f {\cal T}\right) \;,
\end{equation} 
where $P_i$  is the projection operator on the $N_i$-dimensional subspace of eigenstates with quantum numbers $J_i^{\pi_i}, T_i$ ($P_f$ is similarly the projection operator on the corresponding subspace of final states).
The transition intensity is proportional to the square of the matrix element,  and thus has a Porter-Thomas distribution (i.e., Eq. (\ref{chi2-nu}) with $\nu=1$).  

We examined $E2$ and $M1$   $2^{+}, T =1\rightarrow 2^{+}, T=1$
transitions in $A=60$ nuclei.  The $T=1$ states form isobaric 
multiplets with $T_z=0,\pm1$ (i.e., $^{60}$Co, $^{60}$Zn and $^{60}$Cu).  We studied the
statistics of the transition intensities in both $T_z=0$ and $T_z=1$ nuclei.  
Because of the vanishing
of the isovector Clebsch-Gordan coefficient in Eq. (\ref{bel}), the
transitions in the self-conjugate nucleus 
$^{60}$Zn ($T_z=0$) are purely isoscalar.  For each transition operator we
sampled  $66^2 - 66 =4290$ transition matrix elements.  The calculated  distributions
 (histograms) of the $B(E2)$  (left panels) and $B(M1)$ (right panels)
 $2^{+}, T=1\rightarrow 2^{+}, T=1$ transitions are shown in Fig. \ref{fig2}. The top and bottom panels correspond to $T_z=1$ and $T_z=0$ nuclei, respectively.
  All distributions are well described by the Porter-Thomas distribution. Similar results were found by Adams {\it et al}
 \cite{m1} for the $N=Z$  ($T_z=0$) nucleus $^{22}$Na in the $sd$-shell.

When the quantum numbers of the initial and final states are identical, we can also examine the statistics of the diagonal matrix elements $\langle i | {\cal T} | i \rangle$.  Using (\ref{transition-element}) with $i=f$ and the central limit theorem we still find that $\langle i | {\cal T} | i \rangle$ is a Gaussian variable but with a non-zero average 
\begin{equation}\label{diag-mean}
 \overline{\langle i | {\cal T} | i \rangle} = \sum\limits_{m,n} {\cal T}_{m,n} \overline{\psi^\ast_i(m) \psi_i(n)} = {1\over N_i}{\rm Tr}\; \left(P_i {\cal T}\right) \;.
\end{equation}
Using Wick's theorem and the fact that the GOE wave functions $\psi_i(m)$ are real, the variance of a diagonal element is
\begin{equation}
\sigma^2(\langle i | {\cal T} | i \rangle) = \sum\limits_{m,n \atop m',n'} {\cal T}^*_{m,n} {\cal T}_{m',n'}\left[ \overline{\psi_i(m) \psi_i^*(m')} \;\; \overline{\psi_i(n) \psi_i^*(n')} + \overline{\psi_i(m) \psi_i^*(n')} \;\; \overline{\psi_i(n) \psi_i^*(m')} \right] \;.
\end{equation} 
For a real and hermitean operator ${\cal T}$, we find
\begin{equation}\label{diag-variance}
\sigma^2(\langle i | {\cal T} | i \rangle) = {2 \over N_i^2} 
{\rm Tr}\; \left(P_i {\cal T} P_i {\cal T}\right) \;,
\end{equation}
which is a factor of two larger than the variance of an off-diagonal element (see Eq. (\ref{tran-variance}) for the case when the initial and final states span the same subspace of states, i.e.,  $P_i=P_f$). 
 
Since the mean value of $\langle i | {\cal T} | i \rangle$ is non-zero, its square does not follow a Porter-Thomas distribution. However, once the mean value (\ref{diag-mean}) has been subtracted from the diagonal matrix elements, their squares $z$ are predicted to have a Porter-Thomas distribution.  The distributions of the squares of these shifted reduced diagonal matrix elements are shown in Fig. \ref{fig3} for the $M1$ magnetic moments of the $J=2^+,T=1$ states in $A=60$ nuclei.   The values of $z$ are renormalized by dividing out the secular variation with energy of the square of the reduced diagonal matrix element, defined by an equation similar to (\ref{av1}) but with a single Gaussian (since $i=f$).     Considering the small number of data points used to compute the distribution $P(\log_{10}z)$ (there are only 66 diagonal matrix elements), the agreement with a Porter-Thomas distribution (solid lines) is reasonable.

It is of interest to see the effect that the inclusion of diagonal matrix elements has on the distribution of transition intensities. Since the number of diagonal elements is much smaller than the number of transition intensities, we do not expect them to modify the Porter-Thomas statistics of the transition intensities (even though the squares of the unshifted diagonal reduced elements do not follow a Porter-Thomas distribution).  Fig. \ref{fig4} shows the same distributions as in Fig. \ref{fig2} but when the squares of the reduced diagonal elements of the corresponding transition operator are included in the analysis.  The $B(E2)$ distributions as well as the $B(M1)$ distribution in the $T_z=1$ nucleus are all well described  by a Porter-Thomas distribution,  but the $B(M1)$ distribution in the self-conjugate nucleus ($T_z=0$) deviates significantly from the
Porter-Thomas distribution (right bottom panel of Fig. \ref{fig2}). 
The $M1$ transitions in self-conjugate nuclei ($T_z=0$) have the property that they are purely isoscalar. In  $T_z=1$ nuclei  both isoscalar and isovector components contribute
 to the $M1$ transitions, but the isoscalar $M1$ matrix elements
 are much weaker than the isovector $M1$ \cite{Bru}, and the latter dominate the distributions. 
To check that our conclusion does not depend on the size of the model space,  we performed similar calculations in a model space with 6 active
particles, i.e., for $A=62$ nuclei.  Again, we observed that the distributions of the $M1$ transitions in the 
self-conjugate nucleus $T_z=0$ (i.e., $^{62}$Ga) deviate from a Porter-Thomas distribution. 

The deviation from a Porter-Thomas distribution for the isoscalar $M1$ 
can be traced to the diagonal matrix elements.  Although their number is much smaller than the number of off-diagonal elements, they are in magnitude much larger than the off-diagonal elements. Indeed, the isoscalar $M1$ has a particularly simple structure  and can be written (in units of nuclear magneton $\mu_N$) as 
\begin{equation}
{g_p^\ell + g_n^\ell \over 2} {\bf L}+ {g_p^s + g_n^s \over 2} {\bf S} \;,
\end{equation}
  where  ${\bf L}$ and ${\bf S}$ are the total orbital angular momentum and 
spin, respectively, and $g_{p,n}^\ell$, $g_{p,n}^s$ are the proton and neutron
 orbital and spin $g$ factors. Since the transition matrix elements of the 
total angular momentum ${\bf J} = {\bf L} + {\bf S}$ between different states 
vanish, the orbital and spin contributions to the isoscalar $M1$ 
matrix elements always have opposite 
signs. Consequently, the off-diagonal elements are small. However the
interference between the 
orbital and spin contributions to the diagonal matrix elements can be constructive, leading to larger diagonal matrix elements. 
These diagonal matrix elements are related to the magnetic moments, which in 
$T_z=0$ nuclei are given by $J/2 + 0.38 \langle S_z\rangle$ \cite{Bru}. The mean value (\ref{diag-mean}) of the isoscalar magnetic moment is large (relative to a typical off-diagonal element) and leads to the observed deviation from a Porter-Thomas distribution.     
In the case of the isovector $M1$ matrix elements, the relative phase of 
the orbital and spin contributions is more random \cite{Fay}, and the 
corresponding statistics are Porter-Thomas even when diagonal elements are 
included. 

In conclusion, we have studied $\Delta T=0$  $B(E2)$ and $B(M1)$
transition strength distributions in $pf$-shell nuclei with  $A \sim 60$ and found that they agree well with Porter-Thomas statistics.  We have also studied the statistics of diagonal elements (e.g., $M1$ magnetic moments) in these nuclei. The distribution of the squares of these moments is found to be in fair agreement with a Porter-Thomas distribution, but only when the mean value is subtracted from the moments.  In general, the inclusion of off-diagonal elements in the statistics of transition intensities does not affect the observed Porter-Thomas statistics. An exception is the isoscalar $M1$ operator whose diagonal matrix elements are much larger than  its off-diagonal elements.
 
This work was supported
in part by the Department of Energy grant No. DE-FG-0291-ER-40608 and by the Russian
 Foundation for Basic Research, project No.
00-02-17194.

\end{document}